# An Improvement Technique based on Structural Similarity Thresholding for Digital Watermarking

Amin Banitalebi-Dehkordi[1], Mehdi Banitalebi-Dehkordi[2], Jamshid Abouei[3], and Said Nader-Esfahani[4]
[1]Department of Electrical and Computer Engineering, University of British Columbia, Canada
[2]Department of Electrical and Computer Engineering, Ferdowsi University of Mashhad, Iran
[3]Department of Electrical and Computer Engineering, Yazd University, Yazd, Iran
[4]Department of Electrical and Computer Engineering, University of Tehran, Iran
dehkordi@ece.ubc.ca, mahdi_banitalebi@stu.yazd.ac.ir, abouei@yazd.ac.ir, nader@ut.ac.ir

*Abstract*—Digital watermarking is extensively used in ownership authentication and copyright protection. In this paper, we propose an efficient thresholding scheme to improve the watermark embedding procedure in an image. For the proposed algorithm, watermark casting is performed separately in each block of an image, and embedding in each block continues until a certain structural similarity threshold is reached. Numerical evaluations demonstrate that our scheme improves the imperceptibility of the watermark when the capacity remains fix, and at the same time, robustness against attacks is assured. The proposed method is applicable to most image watermarking algorithms. We verify this issue on watermarking schemes in Discrete Cosine Transform (DCT), wavelet, and spatial domain.

## I. Introduction

The rapid growth of the Internet and multimedia technologies has revealed the need for the copyright protection and the proof of the ownership of digital documents [1]. In particular, with images widely available on the Internet, digital watermarking is a common way of identifying images and protecting them from unauthorized usage in online advertisements. In this regard, the most important characteristic of watermark casting is its imperceptibility, where a certain degree of the statistical invisibility of the embedded watermark is required. In addition, in most watermarking techniques, it is desirable to embed messages with appropriate length so that the accurate extraction is assured, and at the same time, the embedded watermark should be perceptually invisible and robust to common signal processing and intentional attacks. Thus, there exists a trade-off between the imperceptibility, capacity and robustness of the watermarking methods [2].

Broadly, watermarking techniques are divided into two categories: i) spatial domain watermarking, and ii) transform domain watermarking. Spatial domain watermarking approaches, where the mark is directly embedded into each pixel of the host image, benefit from the advantages of a low degree of complexity and delay [3]–[5]. In addition, the temporal/spatial localization of the watermark in the spatial domain watermarking schemes is automatically achieved. This permits a better characterization of the distortion introduced by the watermark and reduces the annoying effects. In transform domain watermarking techniques, however, the watermark is inserted into the coefficients of a digital transform of the host asset [6], [7]. For instance, Barni's work [6] which was based on watermarking using the Discrete Cosine Transform (DCT) and the Discrete Fourier Transform (DFT). Of interest is to embed the watermark in the host image using the wavelet domain transform [8]. Authors in [9] utilize a DCT domain watermarking technique for copyright protection of digital images. They propose a watermarking method with dual detection where a pseudo-random sequence of real numbers, as a watermark, is embedded in a selected set of DCT coefficients. Jayalakshmi et al. [10] utilize the multi-resolution dependency of the coefficients of contourlet transform for image watermarking. Generally, transform domain watermarks exhibit a higher robustness to attacks than the spatial domain watermarking schemes. More imperceptibility of the embedded watermark can be achieved by avoiding the changes into low frequency components of the host image. Complexity of this type of embedding might be higher than the complexity of the spatial domain watermarking methods. This issue introduces a trade-off between the robustness and the complexity when using the transform domain watermarking schemes.

Mean Squared Error (*MSE*) is one of the quality measures that has been widely utilized in various watermarking schemes due to its low complexity and ease of usage. However, it is shown that visual systems designed based on the usage of the *MSE* criterion cannot completely track the human visual perception with a high reliability and accuracy [11]. The above challenge motivates us to propose a new technique for improving on digital watermarking algorithms. Our scheme utilizes the structural similarity (*SSIM*) as a quality criterion instead of common measures such as the *MSE* metric. In the proposed scheme, the watermark embedding process is performed on each block of the host image until a certain threshold of the *SSIM* in quality is achieved. The insertion of the watermark sequence in the host image is performed based on an adaptive procedure. The length of the embedded watermark in a loop is increased until a certain threshold of quality is achieved. The stop condition for the watermark insertion process in each block of the image is specified by the *SSIM* threshold. In one hand, each block for the above process should be big enough to allow the *SSIM* efficiently extract local structural similarities, and on the other hand, blocks with larger sizes may increase the complexity.

Numerical results show that the block size of 32×32 yields a lower computational complexity in several algorithms. Our algorithm is numerically tested for various watermarking schemes in the spatial (pixel) domain as well as DCT and wavelet domains. In our simulations, the conventional watermarking methods in the DCT, wavelet and spatial domains are modified so that the capacity is maximized, while the watermarking imperceptibility remains fix. Since our scheme does not change the original way that the watermark is being embedded, it is applicable to many of the watermarking algorithms.

The rest of the paper is organized as follows: In Section II, the structural similarity theory is briefly reviewed. Section III describes the proposed optimization method. Section IV shows the results of numerical experiments. Finally, in Section V, an overview of the results is presented.

## II. STRUCTURAL SIMILARITY

Natural images are highly structured, i.e. there exists a high degree of local correlation between the pixels. To extract the local structural similarity, we follow the primitive formulation of Wang and Bovic in [11], in which the target application is quality assessment of natural images. The original and distorted images, denoted by *x* and *y* respectively, are decomposed to their luminance, contrast, and structure components as follows:

$$l(x,y) = \left( \frac{2\mu_x\mu_y + C_1}{\mu_x^2 + \mu_y^2 + C_1} \right), \quad \text{Luminance} \quad (1)$$

$$c(x,y) = \left( \frac{2\sigma_x\sigma_y + C_2}{\sigma_x^2 + \sigma_y^2 + C_2} \right), \quad \text{Contrast} \quad (2)$$

$$s(x,y) = \left( \frac{\sigma_{xy} + C_3}{\sigma_x\sigma_y + C_3} \right), \quad \text{Structure} \quad (3)$$

where $(\mu_x,\mu_y)$ and $(\sigma_x,\sigma_y)$ represent the local sample means and standard deviations of $(x,y)$, respectively, and $\sigma_{xy}$ denotes the sample cross correlation of *x* and *y* after subtracting their means. In addition, $C_i$, $i = 1, 2, 3$, are small positive constants that stabilize each term, so that near-zero sample means, variances, or correlations do not lead to numerical instability. The local structural similarity (*SSIM*) is formulated as [11]:

$$S(x,y) = l(x,y) \cdot c(x,y) \cdot s(x,y) \quad (4)$$

The *SSIM* metric between the original and the reference image pair $(x,y)$ is calculated by averaging the local *SSIM*, $S(x,y)$, over the image.

## III. IMPROVEMENT TECHNIQUE

As mentioned, regardless of the watermarking algorithm that we would like to improve, the enhancement procedure is performed on each block separately. In each block of the host image, watermark casting is performed via an adaptive procedure exactly the same as the original algorithm that we are modifying, no matter what method is being used for the watermark embedding. To assure the imperceptibility of the mark, the embedding process of the watermark in each block is divided into several steps, and the embedding continues until a certain threshold on the similarity between the primitive block and the current watermarked block is achieved. The above process is performed on all the blocks until the watermarked image is obtained (see Fig. 1). To verify the performance of this approach and compare with the non-modified algorithms, we embed the sequence generated by the concatenation of the sequences that have been embedded in all blocks using the proposed method. Thus, the same sequences

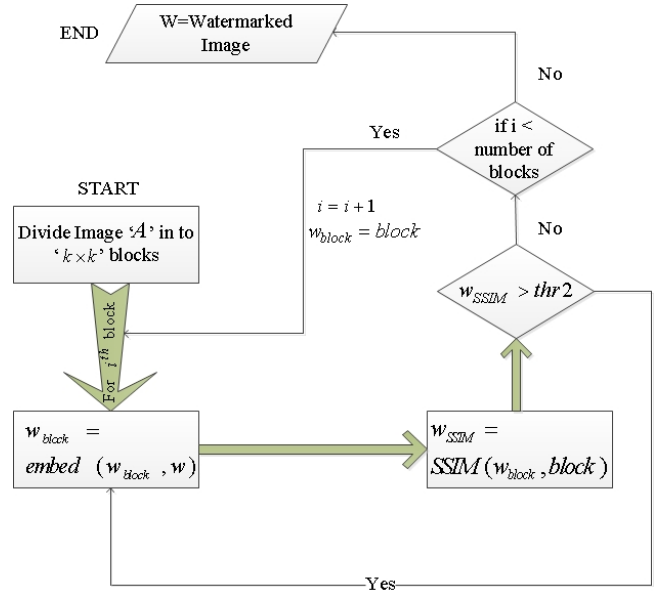

**Fig. 1. Flowchart of the proposed watermarking algorithm**

are embedded using the original and the improved algorithms. In this way, the capacity remains constant and therefore the imperceptibility can be compared for both methods. Since the type of watermark casting (e.g. in spatial, DCT or wavelet domain) is not changed, the robustness will remain almost unchanged. To get more insight into the proposed watermarking algorithm in Fig. 1, we summarize the steps of this figure as follows:

**Step 1**: Divide the asset image ($A$) into $k \times k$ blocks, where a suitable value of $k$ is selected empirically to reduce the computational complexity. For our algorithm, an efficient value of $k$ is 32 for the sake of the complexity reduction.

**Step 2**: Set $i = 1$, name the $i^{th}$ block of $A$ as "*block*", and define $w_{block}$ = block. $w_{block}$ will be the $i^{th}$ block after the watermark insertion.

**Step 3**: For a specific watermarking algorithm, embed the watermark sequence in "*block*" to create $w_{block}$.

**Step 4**: Compute the *SSIM* between "block" and $w_{block}$. If this similarity is greater than a pre-specified threshold *thr2*, then go to Step 3 and embed a new watermark sequence in "block". The new watermark has a bigger length when compared to the previous watermark. Watermark is usually a sequence of random bits. The new mark is the concatenation of the previous mark and $L$ new random bits, where $L$ is chosen corresponding to the watermark casting algorithm. If the mark is an image, the new mark can be a higher resolution image. For the Least Significant Bit (LSB) watermarking, the previous mark would be the $(L-1)^{th}$ left-sided bits of the mark image, while the new mark would be the $L^{th}$ left-sided bit planes.

**Step 5**: Repeat Step 4 until the *SSIM* becomes less than *thr2*.

**Step 6**: Once the *SSIM* value is less than *thr2*, the current $w_{block}$ is the ultimate marked block of the watermarked image. Save this block, set $i = i + 1$ and reset $w_{block}$ to the next block. Then go to Step 4 and perform the same process for the next block of $A$.

Note that the iterative block embedding procedure is repeated for all of the blocks of the original image. Then, the output of the algorithm would be the watermarked image $W$. The proposed algorithm leads to an optimum watermark embedding in the sense that by adjusting the threshold value, one can directly tune the imperceptibility and the capacity. According to Fig. 1, the structural similarity based method offers an adaptive embedding procedure that embeds the watermark sequence until a certain threshold for the imperceptibility is achieved. Clearly, less threshold values results in a higher capacity but less invisibility. Another advantage of the proposed method is that the scheme is robust against most of the intentional and unintentional attacks. As we will show in

Section IV, for most of the watermark casting algorithms our scheme can extract the true watermark sequence with a higher accuracy than each original algorithm.

IV. NUMERICAL RESULTS

This section presents some numerical evaluations for the proposed watermarking algorithm, where we use the Lena image as asset. The proposed method has been examined by modifying four watermarking schemes described briefly in the subsequent sections. Afterwards we will state our innovation. In our simulation, the threshold *thr2* for *SSIM* is empirically set to 0.8.

*A. Pixel Domain Embedding: LSB Watermarking*

For the original LSB algorithm, watermark is an image which is embedded in another host picture. For each pixel, the $N$ left-sided bit planes of the mark image are replaced with the $N$ right-sided bits of the asset. In this way, most significant bits (i.e. important information) of the watermark image are replaced with the least significant bits (or details) of the host image. Within an adaptive process, in each time slot, a bit plane is added to the embedding bits of the watermark image block. The embedding process is stopped when the threshold is reached or all bits are embedded. As previously mentioned, the embedding process is performed adaptively in each block.

*B. DCT Domain Embedding: Barni's Method [12]*

For the original algorithm described in [12], the watermark is a sequence of random bits placed in an additive form to the DCT coefficients of the host image as follows:

$$t'_{L+i} = t_{L+i} + \alpha |t_{L+i}| x_i \;,\; i=1,2,...,M \quad (5)$$

where $X = \{x_1, x_2, ..., x_M\}$ is the embedding sequence, $t_j$ and $t'_j$ are the DCT coefficients of the host and the watermarked images, respectively, and $\alpha$ is a parameter to adjust the watermark power. For a better comparison with our method, the visual masking effect (i.e., the extra stage for improving the imperceptibility) is turned off in both the original and the proposed block-wise algorithm. The same steps are done for the proposed algorithm; the only difference is that the embedding process is performed on each single block separately. In comparison to the original method, the watermarking sequence length is much smaller for each block, but larger for the entire image. The value of $\alpha$, and the length of the watermark sequence are increased iteratively in each loop cycle until the similarity threshold is reached (*SSIM* between the watermarked and the original blocks meets the threshold).

*C. Wavelet Domain Embedding: Ahuja's Method [13]*

For the original algorithm in [13], the watermark casting is performed in the wavelet domain. Authors in [13] utilize the wavelet decomposition property to divide the host image to different frequency resolutions. Similar to the embedding process in the DCT domain, the watermark sequence is adaptively embedded in the middle frequency wavelet coefficients. For the proposed algorithm, casting of the mark is done in each block separately. The embedding process is done iteratively. Insertion of the watermark in each of the iterations is performed on 2000 wavelet coefficients more than the previous iteration. The embedding process continues until the *SSIM* threshold is reached.

*D. Wavelet Domain Embedding: CDMA-Based Method [14]*

For the original algorithm, the watermark casting is based on the Code Division Multiple Access (CDMA) and multilevel coding [14]. The bits of the watermark are grouped together and for each sequence a different modulation coefficient is used. Different watermark messages are hidden in the same transform coefficients of the cover image using orthogonal or semi-orthogonal codes [14]. The proposed algorithm is the same as the original one. However, it is performed for each individual block separately. Size of the mark increases during each iteration until the *SSIM* threshold is reached.

Fig. 2 demonstrates the original and the watermarked images for all four algorithms mentioned above. This figure also shows the effect implied by our method on the images. As observed in Fig. 2, the proposed method improves the imperceptibility property of the watermark while the capacity remains constant in both original watermarking method and the proposed approach. Table I illustrates the statistical invisibility of the watermark. Similarity in terms of the *MSE* and the *SSIM* between the original and the watermarked images is given in Table I. It is revealed that the similarity and therefore the imperceptibility are maintained more in the proposed algorithm than in the original methods. In some cases, the difference is salient.

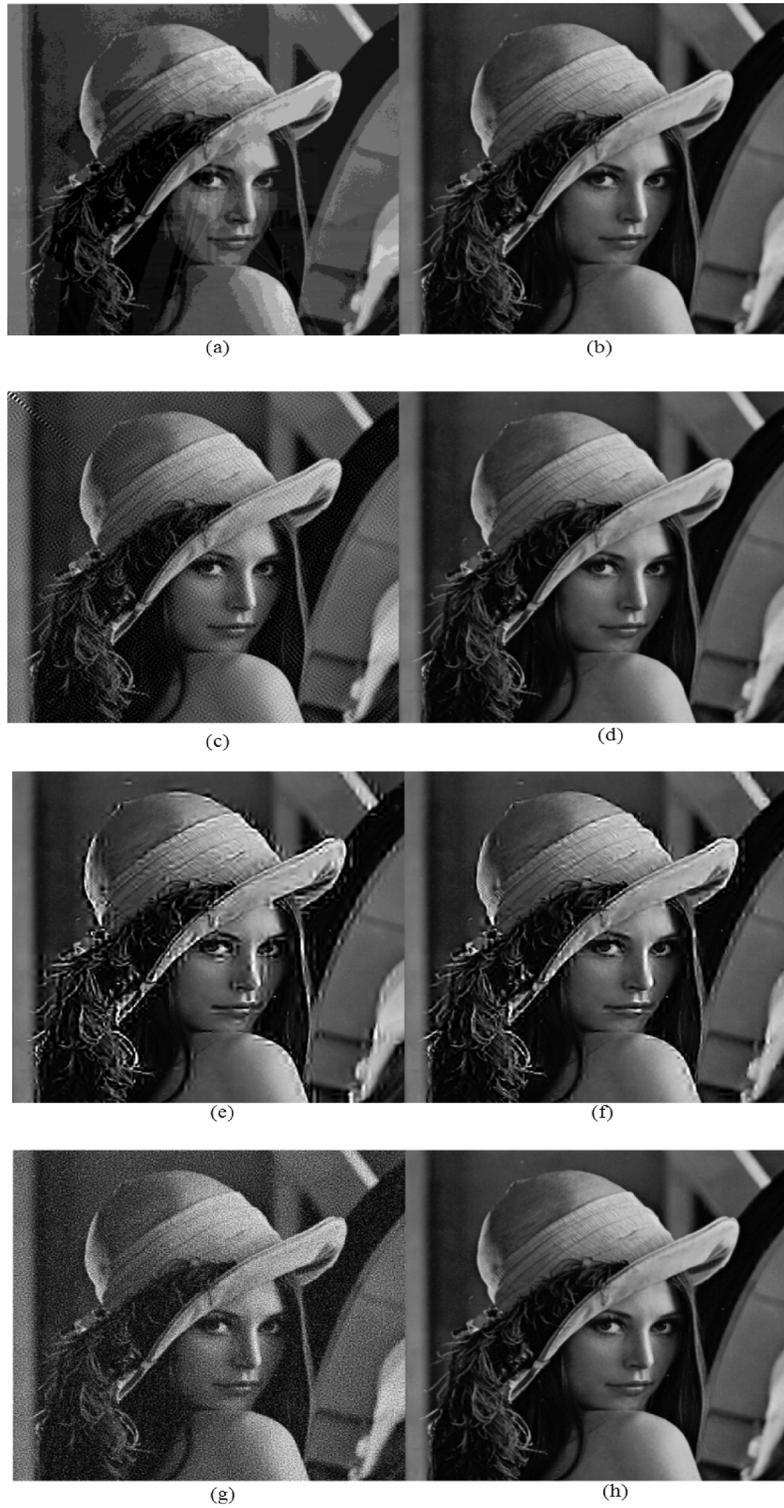

Fig. 2. LSB algorithm: original (a), proposed (b), DCT-based algorithm: original (c), proposed (d), wavelet-based algorithm: original (e), proposed (f), CDMA-based algorithm: original (g), proposed (h)

The complexity of the proposed method highly depends on the complexity of the original method that our approach is modifying. As Table I shows, the complexity for the LSB watermarking is highly reduced when it is improved by our method. However, the proposed method increases the complexity for the DCT-based and the wavelet-based watermarking methods. For the CDMA-based scheme, the complexity is not changed. Taking the above considerations into account and this fact that our approach performs the same trend as the original approach, we conclude that the complexity of the proposed method depends on the original watermarking scheme.

Fig. 3 demonstrates the response of watermark detectors to attacked watermarked images. After the original image is watermarked by the proposed methods (i.e. modified version of the four algorithms), the following three types of attacks are applied to them: Gaussian noise, Low Pass Filtering (LPF), and JPEG compression. Attacks have the same parameters and conditions for all images. The watermark is considered as a sequence of bits, even for the LSB watermarking. 300 random sequences are generated where three of them are replaced with the extracted sequences from the attacked images. Fig. 3 shows the correlations between the true watermark sequence and the above 300 sequences. It is revealed from this figure that the proposed method is robust in almost all attacks. This robustness is followed from the robustness of the original algorithms, in particular for the DCT-based and the wavelet-based algorithms which are robust against attacks.

TABLE I. WATERMARK IMPERCEPTIBILITY AND ALGORITHM COMPLEXITY FOR THE ORIGINAL AND PROPOSED METHODS

|  | *MSE* | | *SSIM* | | Algorithm Complexity: Simulation Time (Sec) | |
| --- | --- | --- | --- | --- | --- | --- |
|  | *Original* | *Proposed* | *Original* | *Proposed* | *Original* | *Proposed* |
| LSB | 11.28 | 0.03 | 0.72 | 0.98 | 99 | 4 |
| DCT-based [12] | 50.64 | 0.5 | 0.5 | 0.96 | 2 | 22 |
| Wavelet-based [13] | 26.03 | 16.48 | 0.75 | 0.95 | 4 | 35 |
| CDMA-Wavelet [14] | 71.59 | 9.13 | 0.42 | 0.97 | 228 | 231 |

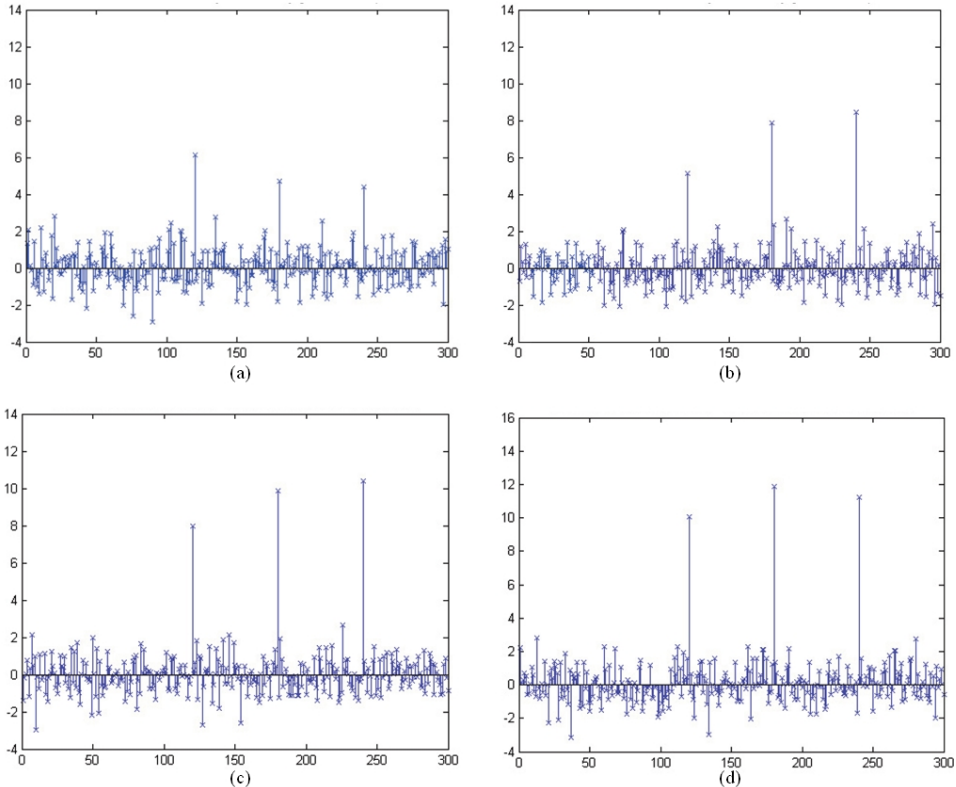

**Fig. 3. Watermark detection performance for the proposed method. High bars that stand out in each figure represent Gaussian noise, low pass filtering (LPF), and JPEG compression, from left to right respectively: (a) LSB watermarking, (b) DCT-based watermarking, (c) wavelet-based watermarking, and (d) CDMA-based watermarking**

*E. Sensitivity of the proposed method to the block sizes*

It was mentioned in Section III that the blocks for watermark embedding are of $k \times k$ size. We selected $k$ equal to 32 for an efficient trade-off between the performance of the proposed method and computational complexity. In this sub-section, we study the effect of the block size on the embedding performance as well as the computational complexity. To this end, the performance and complexity are evaluated at various block sizes. The same parameters as Table I are used so that the *SSIM* values for the original algorithms are 0.72, 0.5, 0.75, and 0.42 for the LSB, DCT-based [12], Wavelet-based [13], and CDMA-Wavelet [14] methods, respectively. Fig. 4 illustrates the trade-off between the performance, complexity, and the block size. It is observed from Fig. 4 that changes in block size result in smooth variations in the watermark imperceptibility and complexity of different algorithms. We chose a block size of 32, however, any other block size can be chosen.

*F. Sensitivity of the proposed method to the threshold value*

The block embedding threshold value (*thr2*) specifies the length of the watermark which is embedded in each block. The lower the threshold is, the more number of bits will be embedded in each block and therefore a higher capacity is achieved. However, a lower threshold results in a more perceptible mark in the host image. To find a suitable threshold value, we sweep the threshold value from 0.6 to 0.9 and evaluated the *SSIM* and *MSE* values for the marked image compared to the original one. Results are illustrated in Table II. It is observed from Table II that the *SSIM* and *MSE* values vary smoothly when the threshold changes. This shows the robustness of the proposed method to the threshold value. We chose a threshold value of 0.8. However, this value can be changed according to the desired amounts of imperceptibility and capacity.

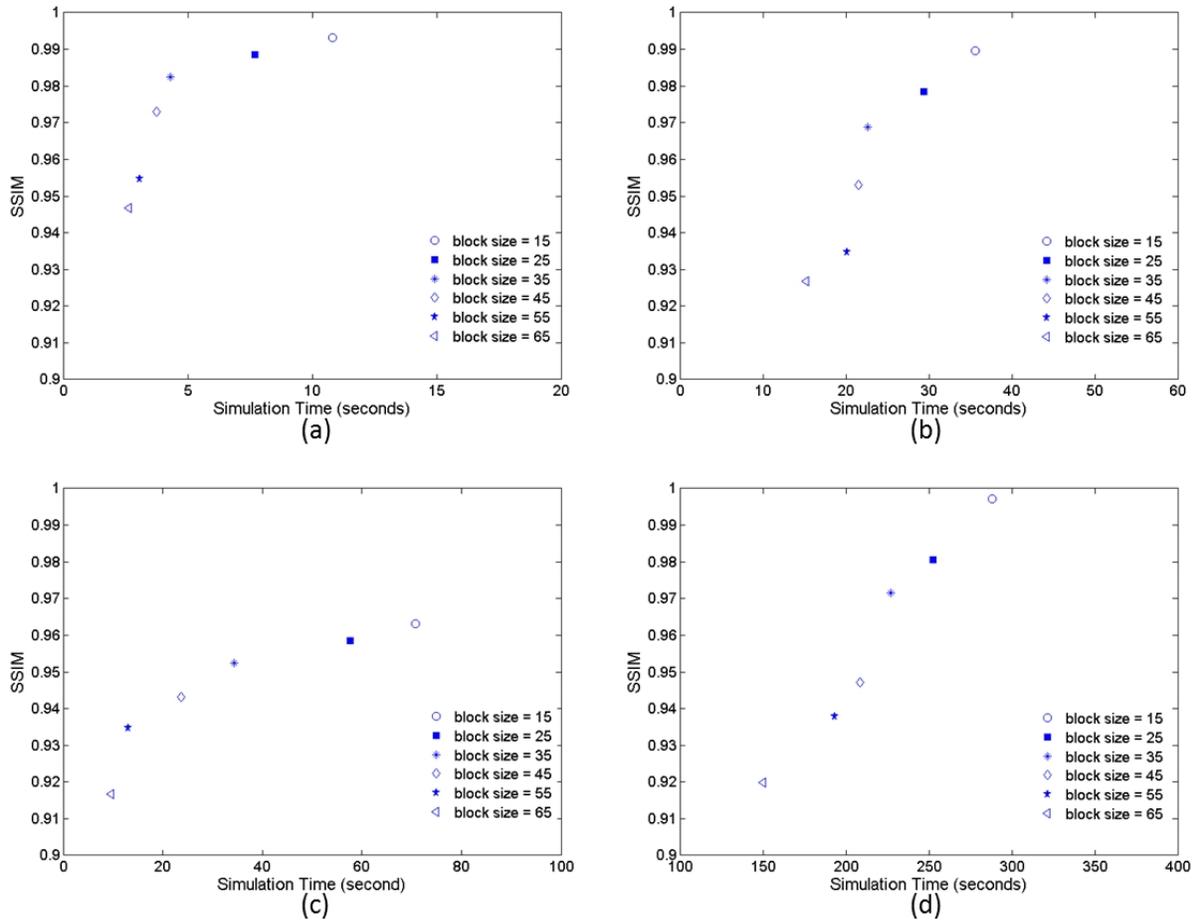

Fig. 4. Imperceptibility and complexity for various block sizes: (a) LSB watermarking, (b) DCT-based watermarking, (c) wavelet-based watermarking, and (d) CDMA-based watermarking

TABLE II. WATERMARK IMPERCEPTIBILITY AND ALGORITHM COMPLEXITY FOR VARIOUS THRESHOLD ($Thr2$) VALUES

|  | $Thr2=0.6$ | | $Thr2=0.7$ | | $Thr2=0.8$ | | $Thr2=0.9$ | |
| --- | --- | --- | --- | --- | --- | --- | --- | --- |
|  | *MSE* | *SSIM* | *MSE* | *SSIM* | *MSE* | *SSIM* | *MSE* | *SSIM* |
| LSB | 28.22 | 0.91 | 12.39 | 0.94 | 0.03 | 0.98 | 0.01 | 1 |
| DCT-based [12] | 37.78 | 0.88 | 14.18 | 0.93 | 0.5 | 0.96 | 0.13 | 1 |
| Wavelet-based [13] | 39.39 | 0.87 | 23.96 | 0.92 | 16.48 | 0.95 | 1.43 | 0.99 |
| CDMA-Wavelet [14] | 36.36 | 0.89 | 14.07 | 0.94 | 9.13 | 0.97 | 0.71 | 1 |

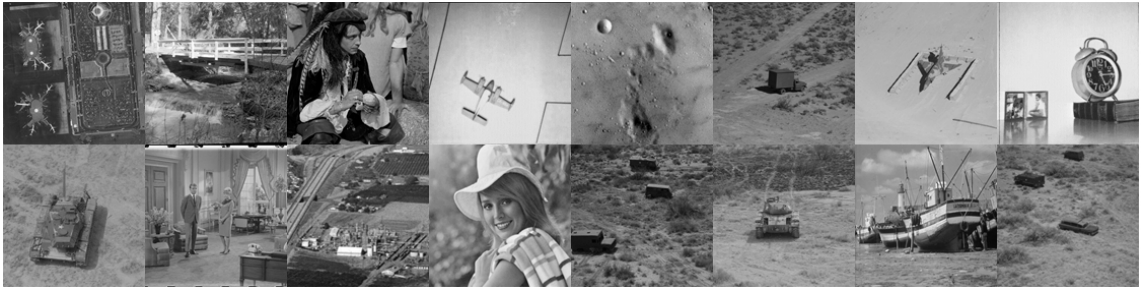

**Fig. 5. Snapshots of the incorporated image dataset**

*G. Validation of the proposed method over a dataset of images*

We validate the performance of the proposed method for a dataset of images to verify the generalization of our approach. Fig. 5 shows snapshots of the incorporated image set. Using this set of images and the same threshold parameters used in Section IV, the proposed method resulted in an average of 25 %, 44 %, 23 %, and 53 % improvements in visual quality (*SSIM*) for LSB, DCT-based [12], Wavelet-based [13], and CDMA-based [14] watermarking methods, respectively.

## V. CONCLUSION

In this paper, we proposed a new method to improve the watermark embedding procedure. We suggested that many embedding algorithms can be performed via each block separately and for all the blocks. This can be performed during an adaptive iterative process. The stop condition for this procedure is a threshold on the structural similarity that is compatible with the human visual system. Numerical results showed that the proposed method can improve the imperceptibility of the watermark for various examined watermark casting algorithms. It was seen that the capacity remains constant and the complexity varies among different marking algorithms. After intentional attacks such as Gaussian noise, low pass filtering, and JPEG compression, the watermark detector showed noticeable correlation between the extracted watermarks and the true watermarks which verifies that the proposed algorithm is robust against these attacks.

REFERENCES

[1] I. J. Cox, M. L. Miller, J. A. Bloom, J. Fridrich, and T. Kalker, Digital Watermarking and Steganography, Morgan Kaufmann, 2nd ed., 2008.

[2] X. Gong and H.-M. Lu, "Towards fast and robust watermarking scheme for H.264 video," in Proc. 10th IEEE International Symposium on Multimedia (ISM), Dec. 2008, pp. 649–653.

[3] S. S. Bedi and S. Verma, "A design of secure watermarking scheme for images in spatial domain," in Proc. Annual IEEE India Conference, Sept. 2006.

[4] H. H. Larijani and G. R. Rad, "A new spatial domain algorithm for gray scale images watermarking," in Proc. International Conference on Computer and Communication Engineering (ICCCE), May 2008, pp. 157–161.

[5] V. Licks, F. Ourique, R. Jordan, and F. Perez-Gonzalez, "The effect of the random jitter attack on the bit error rate performance of spatial domain image watermarking," in Proc. International Conference on Image Porcessing (ICIP), Sept. 2003, pp. 455–458.

[6] M. Barni, F. Bartolini, A. De Rosa, and A. Piva, "Capacity of the watermarking channel: how many bits can be hidden within a digital image," in Proc. Security and Watermarking of Multimedia Contents, Procedeeings of SPIE, Jan. 1999, vol. 3654, pp. 437–448.

[7] A. Banitalebi, S. Nader-Esfahani, and A. Nasiri Avanaki, "Robust LSB watermarking optimized for local structural similarity," in Proc. 19th Iranian Conference on Electrical Engineering (ICEE), May. 2011.

[8] Y. Yusof and O. O. Khalifa, "Imperceptibility and robustness analysis of DWT-based digital image watermarking," *in Proc. IEEE International Conference on Computer and Communication Engineering (ICCCE)*, May 2008, pp. 1325–1330.

[9] G. P. Betancourth, A. Haggag, M. Ghoneim, T. Yahagi, and J. Lu, "Robust watermarking in the DCT domain using dual detection," *in Proc. IEEE International Symposium on Industrial Electronics*, July 2006, pp. 579–584.

[10] M. Jayalakshmi, S. N. Merchant, and U. B. Desai, "Significant pixel watermarking in contourlet domain," *in Proc. IET International Conference on Visual Information Engineering (VIE'06)*, Sept. 2006, pp. 416–421.

[11] Z. Wang and A. C. Bovik, "Mean squared error: love it or leave it? a new look at signal fidelity measures," *IEEE Signal Processing Magazine*, vol. 26, no. 1, pp. 98–117, Jan. 2009.

[12] A. Piva, M. Barni, F. Bartolini, and V. Cappellini, "DCT-based watermark recovering without resorting to the uncorrupted original image," *in Proc. International Conference on Image Processing (ICIP)*, Oct. 1997, pp. 520–523.

[13] R. Dugad, K. Ratakonda, and N. Ahuja, "A new wavelet based scheme for watermarking images," *in Proc. International Conference on Image Processing (ICIP)*, Oct. 1998, pp. 419–423.

[14] S. P. Maity and M. K. Kundu, "A blind CDMA image watermarking scheme in wavelet domain," *in Proc. International Conference on Image Processing (ICIP)*, Oct. 2004, pp. 2633–2636.